\documentclass[11pt]{article}
\usepackage{jheppub}
\usepackage{amsfonts,ulem,bbold}
\usepackage{amsmath,amssymb, braket}
\newcommand{\te}{\tilde\epsilon}
\newcommand{\cN}{{\cal N}}
\newcommand{\mpl}{M_{\rm Pl}}
\newcommand{\be}{\begin{equation}} 
\newcommand{\ee}{\end{equation}}
\newcommand{\ba}{\begin{array}} 
\newcommand{\ea}{\end{array}}
\newcommand{\bea}{\begin{eqnarray}} 
\newcommand{\eea}{\end{eqnarray}}
\newcommand{\beqn}{\begin{eqnarray}}
\newcommand{\eeqn}{\end{eqnarray}} 
\newcommand{\p}{\partial}
\newcommand{\nn}{\nonumber} 

\newcommand{\ds}{\displaystyle}

\def\mpl{M_{\mathrm{Pl}}}

\newcommand{\md}{\mathrm{d}}

\title{Metric Formulation of Ghost-Free Multivielbein Theory} 
\author{S.F.~Hassan,} 
\author{A.~Schmidt-May} 
\author{and M.~von~Strauss} 
\affiliation{Department of Physics \& The Oskar
  Klein Centre,\\ 
Stockholm University, AlbaNova University Centre, SE-106 91
Stockholm, Sweden}  
\emailAdd{fawad@fysik.su.se}
\emailAdd{angnis.schmidt-may@fysik.su.se} 
\emailAdd{mvs@fysik.su.se}
\abstract{We formulate the recently proposed ghost-free theory of
  multiple interacting vielbeins in terms of their corresponding
  metrics. This is achieved by reintroducing all local Lorentz
  invariances broken by the multivielbein interaction potential which,
  in turn, allows us to explicitly separate the gauge degrees of
  freedom in the vielbeins from the components of the metrics by an
  appropriate gauge choice. We argue that the gauge choice does not
  spoil the no-ghost proof of the multivielbein theory, hence the
  multimetric theory is ghost-free. We further show the on-shell
  equivalence of the metric and vielbein descriptions, first in
  general and thereafter in two illustrative examples.}
\preprint{}
\toccontinuoustrue
\begin{document} 
\maketitle
\flushbottom
\section{Introduction and summary}
Theories of massive and multiple interacting spin-2 fields are often
discussed in terms of theories of massive gravity and bimetric or
multimetric gravity. Until recently constructing theories of this
type, that were at least classically consistent, had remained an
unsolved problem. This was due to the presence of Boulware-Deser
ghosts \cite{Boulware:1972zf, Boulware:1973my} that generically arise
in massive spin-2 theories. The situation drastically changed since
the massive gravity proposal of \cite{deRham:2010ik, deRham:2010kj},
that could describe a massive spin-2 field in flat spacetime. That
this model avoided the BD ghost at the nonlinear level was proven in
\cite{Hassan:2011hr, Hassan:2011tf, arXiv:1111.2070}, which also
extended it to a massive spin-2 field in curved spacetime.

Subsequently, a ghost-free theory of two interacting spin-2 fields was
constructed in \cite{Hassan:2011zd} as a bimetric theory (for earlier
work on bimetric theories see
\cite{Rosen:1940zz,Rosen:1975kk,Isham:1971gm,Salam:1976as,
  Boulanger:2000rq,Damour:2002ws,Damour:2002wu,ArkaniHamed:2002sp}).
As a generalization of the bimetric theory, in a remarkable recent
work an interacting multivielbein theory of ${\cal N}$ vielbeins was
shown to be ghost-free \cite{Hinterbichler:2012cn}. For ${\cal N}=2$
it was possible to express the interactions in terms of metrics,
reobtaining the bimetric theory. For earlier attempts at vielbein
formulations see
\cite{CSS,Nibbelink:2006sz,Chamseddine:2011mu,Volkov:2011an}. 

A difficulty encountered in \cite{Hinterbichler:2012cn} was in finding
an expression for the multivielbein interactions in terms of metrics
for ${\cal N}>2$. If the multivielbein theory is regarded as a theory
of interacting spin-2 fields, then finding such a metric description
is desirable since in setups with general covariance, metrics provide
a minimal description of spin-2 fields.

To summarize, in this paper we show that a gauge invariant
generalization of the multivielbein theory can indeed be written as a
ghost-free theory of $\cal N$ interacting metrics. In
\cite{Hinterbichler:2012cn}, the authors consider a set of interaction
terms involving $\cN$ vielbeins with a single local Lorentz
invariance. We reintroduce the $\cN-1$ broken local Lorentz
invariances, now allowing for different gauge fixings, and argue that
the no-ghost proof of \cite{Hinterbichler:2012cn} is still valid. This
enables us to express the interaction potential in terms of $\cal N$
metrics as well as ${\cal N}-1$ non-dynamical antisymmetric tensor
fields. The latter are determined by their equations of motion which,
as we demonstrate, are equivalent to the constraint equations that
arise in the vielbein formulation.

\section{Review of multivielbein theory}

In this section we describe the multivielbein theory that was recently
shown to be free of the Boulware-Deser ghosts
\cite{Boulware:1973my,Boulware:1972zf} in \cite{Hinterbichler:2012cn}.  
We also discuss the generality of the no-ghost proof of
\cite{Hinterbichler:2012cn} which is essential for the consistency of
the multimetric formulation of the theory. 

{\it The action}\,: The multimetric theory is constructed in terms of
$\cN$ vielbeins ${E^a}_{\mu}(I)$, with associated metrics, 
\be
\label{metrics} 
g_{\mu\nu}(I)=E_{~\mu}^{a}(I)\,\eta_{ab}\,E^b_{~\nu}(I)\,,  
\ee 
where $I=1,\cdots,\cN$. The kinetic part of the action is a sum of
Einstein-Hilbert actions for the individual metrics,  
\be
\label{KE}
\sum_{I=1}^{\cN}\mpl^{d-2}(I)\int d^dx \sqrt{-\det{g(I)}}~R(g(I))\,.
\ee 
The interaction part of the action is the mass potential,
\be
\frac{m^2}{4}\int d^dx\, U = \sum_{\{I\}}
\frac{m^2}{4}\int d^dx\,T^{I_1\cdots I_\md}\,U_{I_1\cdots I_\md } 
\label{U}
\ee
where the $T^{I_1\cdots I_d}$ are constant coefficients of mass
dimension $d-2$ and the functions $U_{I_1\cdots I_d }$ are
given entirely in terms of the vielbeins \footnote{For $\cN =2$, this
  type of structure appeared in \cite{CSS} in the context of
  supersymmetric bimetric gravity. Massive gravity in terms of
  vielbeins was also discussed in \cite{Nibbelink:2006sz}.}, 
\be
U_{I_1\cdots I_\md}=
\te^{\mu_1\cdots \mu_\md}
\te_{a_1\cdots a_\md}\,E^{a_1}_{~\mu_1}(I_1)\cdots E^{a_\md}_{~\mu_\md}(I_{\md})
\label{mass}
\ee
Here, both $\te^{\mu_1\cdots\mu_\md}$ and $\te_{a_1\cdots a_\md}$ are
tensor densities which means that they are, effectively, invariant
under general coordinate and Lorentz transformations respectively. 
Then (\ref{mass}) transforms in the same way as the usual volume form
$\sqrt{-\det g}$ and the action is invariant. 

{\it Symmetries}\,: The Einstein-Hilbert actions (\ref{KE}) are
invariant under independent local Lorentz transformations of the $\cN$
vielbeins $E_{~\mu}^{a}(I)$. But in writing the interaction terms
(\ref{mass}), the Lorentz frames of all vielbeins have been identified
with each other and with that of $\te_{a_1\cdots a_\md}$ so that these
terms are invariant under a single local Lorentz group that acts on
all vielbeins in the same way. The remaining $\cN -1$ broken Lorentz
groups have, in fact, been used to identify the frames to get to
(\ref{mass}), as will become clearer in the next section. Hence, these
$\cN-1$ broken groups are no longer available to gauge away
$(\cN-1)\times d(d-1)/2$ components of the vielbeins, or to rotate
them to other frames. In this sense, the $E^a_{~\mu}(I)$ in
(\ref{mass}) appear in given Lorentz frames, each with $d^2$
independent components, modulo the overall local Lorentz
transformations.

{\it Equations of motion}\,: In the multivielbein theory, these are
$\delta S/\delta E^a_{~\mu}(I)=0$, that is $d^2$ equations for each
$I$. Derivatives of $E^a_{~\mu}(I)$ or $g_{\mu\nu}(I)$ appear only
through the Einstein tensor and hence are contained in the symmetric
combinations,    
\be
\frac{\delta S}{\delta {E^a}_\mu(I)}{E^c}_\mu(I)\eta_{cb}+
\frac{\delta S}{\delta {E^b}_\mu(I)}{E^c}_\mu(I)\eta_{ca}=0\,.
\label{Eeom-sym}
\ee
The remaining antisymmetric combinations do not contain derivatives
and thus give $d(d-1)/2$ constraints equations 
\cite{Hinterbichler:2012cn},   
\be
\frac{\delta U}{\delta {E^a}_\mu(I)}{E^c}_\mu(I)\eta_{cb}-
\frac{\delta U}{\delta {E^b}_\mu(I)}{E^c}_\mu(I)\eta_{ca}=0\,.
\label{Eeom-antisym}
\ee 
This is the correct number of equations to reduce the number of
independent components of each vielbein from $d^2$ to $d(d+1)/2$, 
equal to the number of independent components of the corresponding
metric. In practice, non-trivial equations (\ref{Eeom-antisym}) exist
only for $\cN -1$ of the vielbeins (due to the overall Lorentz
invariance of the action), while the last vielbein can be reduced to
$d(d+1)/2$ components using this overall local Lorentz invariance
\cite{Hinterbichler:2012cn}. The surviving $\cN d(d+1)/2$ components
are then governed by (\ref{Eeom-sym}).  

{\it Attempt at a multimetric description}\,: Based on the above
discussion, it is natural to conjecture that solving the constraint
equations (\ref{Eeom-antisym}) may hopefully lead to expressions for
the vielbeins in terms of the metrics and thus enable us to express
the multivielbein interactions (\ref{mass}) as multimetric
interactions.  However, this scheme is difficult to realize beyond
$\cN =2$. Indeed in \cite{Hinterbichler:2012cn} it was found that for
$\cN =2$, the constraints (\ref{mass}) are solved by a certain
condition on the two vielbeins that also expresses them in terms of
two metrics, thus recovering the bimetric theory of \cite{Hassan:2011zd}.
However, \cite{Hinterbichler:2012cn} also found that for $\cN >2$
(except for cases that are trivial extensions of $\cN=2$), the
constraints (\ref{mass}) depend on the free parameters of the theory
and it is not easy to find solutions that lead to generic
expressions for the vielbeins in terms of the metrics. 

While in many ways, working with the multivielbein theory may be
easier than working with the corresponding multimetric theory, the
motivation for expressing the multivielbein action in terms metrics is
the final aim of using this as a theory of interacting spin-2
fields. In setups with general covariance, we are familiar with
describing spin-2 fields in terms of symmetric rank-2 tensors. In this
sense, spin-2 fields are more closely related to metrics than to
vielbeins where one has to solve more equations to get to the spin-2
content. Hence, as a theory of spin-2 fields, it is desirable that the
multivielbein interactions are also expressible in terms of the
metrics, or rank-2 symmetric tensors in general. An expression for the
multivielbein interactions in terms of metrics is obtained in the next
section.

{\it Generality of the no-ghost proof}\,: In
\cite{Hinterbichler:2012cn}, Hinterbichler and Rosen showed that the
above multivielbein theory has the right number of constraints to
eliminate the Boulware-Deser ghosts. Here we review the main aspect of
the ghost analysis, in particular emphasizing that the no-ghost proof
of \cite{Hinterbichler:2012cn} is more general than the setup
considered there explicitly. This is required for the consistency of
the multimetric formulation found the in next section.

To analyze the ghost content at the nonlinear level one has to work in
the ADM formulation and write the metrics $g_{\mu\nu}(I)$ in a $1+(d-1)$
decomposition in terms of the lapses $N(I)$, shifts $N_i(I)$ and
spatial metrics $g_{ij}(I)$ \cite{Arnowitt:1962hi}. Such metrics are
obtained from Lorentz transformations of constrained vielbeins and
\cite{Hinterbichler:2012cn} chooses the parameterization,
\be 
E^{a}_{~\mu}(I)= \hat\Lambda^a_{~b}(\alpha_r;I)\hat E^a_{~\mu}\,,
\qquad \hat E^a_{~\mu}=
\left(\ba{ccc}N(I) && 0  \\ 
e^{\hat b}_{~j}(I)\,N^j(I) && e^{\hat b}_{~i}(I)\ea\right)\,.
\label{Eadm}
\ee 
$E^a_{~\mu}$ has $d^2$ parameters: the $\hat\Lambda$, satisfying
$\hat\Lambda^T\eta\hat\Lambda=\eta$, depend on $d(d-1)/2$
parameters $\alpha_r$ and, for definiteness, we take the constrained
vielbeins $\hat E$ to depend only on the $d(d+1)/2$ parameters that
also appear in the metric.

The no-ghost proof \cite{Hinterbichler:2012cn} rests on the
possibility that all $E(I)$ can be written as in (\ref{Eadm}) with
Lorentz matrices $\hat\Lambda(\alpha_r;I)$ that do not depend on the
$N(I)$ and $N_i(I)$. Then only the first column of $E^a_{~\mu}$
contains linear combinations of $N$ and $N_i$ and, due to the
antisymmetry of (\ref{mass}), the potential $U$ is linear in these
variables. Hence, $N(I)$ and $N_i(I)$ are Lagrange multipliers and
their equations of motion contain the primary constraints that
eliminate the ghosts (secondary constraints that remove the momentum 
conjugate to the ghost are also expected to arise as in
\cite{arXiv:1111.2070,Hassan:2012qv}).

To obtain a multimetric form for (\ref{mass}), one needs to go through
vielbeins that are more general than (\ref{Eadm}). In the ADM
formalism, a generic vielbein can always be put in the form  
\be 
E^{a}_{~\mu}(I)=\tilde\Lambda^a_{~b}(\hat E,\alpha_r;I)
\hat E^a_{~\mu}(I)\,,
\label{Eadm-gen}
\ee 
where $\hat E$ has the same form as in (\ref{Eadm}) with $d(d+1)/2$
parameters, but now the Lorentz matrix $\tilde\Lambda$ may
depend on the parameters of $\hat E $ (including the $N$ and $N_i$)
along with $d(d-1)/2$ other independent parameters $\alpha_r$. This
cannot be converted to the form (\ref{Eadm}) by available symmetries
and hence does not satisfy the no-ghost criterion straightforwardly. 

However, it is easy to argue that the no-ghost proof of
\cite{Hinterbichler:2012cn} also applies to (\ref{Eadm-gen}) as is 
best seen in the Cayley parameterization of the Lorentz matrix
$\tilde\Lambda$ (see equation (\ref{Cayley})). This makes it very
explicit that the $d^2$ components of $\tilde\Lambda$ depend on the
components of $\hat E$ and the $\alpha_r$ only through $d(d-1)/2$
functions $\tilde A_r$ (which in this parameterization are arranged as 
components of an antisymmetric tensor $\tilde A_{ab}$). Explicitly
(suppressing the index $I$), 
\be
\tilde\Lambda^a_{~b}=\tilde\Lambda^a_{~b}(\tilde A_r (\hat E,\alpha)) 
\ee
Since $E^a_{~\mu}$ depends on $d^2$ independent parameters, the
relation between the $\tilde A_r$ and the functions $\alpha_r$
is invertible. This insures that if we now make a field redefinition
from the fields $\alpha_r$ to the fields $\tilde A_r$ and
treat the new fields as independent of $N$ and $N_i$, then the set
of ($\hat E, \alpha_r$) equations of motion are equivalent to the set
of ($\hat E,\tilde A_r$) equations of motion, as can be
explicitly checked.\footnote{This is the analogue of ``field
  redefinitions'' encountered in the no-ghost proofs of massive
  gravity and bimetric gravity
  \cite{deRham:2010kj,Hassan:2011hr,Hassan:2011tf,Hassan:2011zd}} 
In terms of the redefined fields, the $E^a_{~\mu}$
of (\ref{Eadm-gen}) satisfy the same no-ghost criteria as vielbeins
in (\ref{Eadm}) and the no-ghost proof of \cite{Hinterbichler:2012cn}
still applies. This fact is important for the absence of ghosts in the
multimetric form of the action obtained in the next section.

\section{Multimetric action from the multivielbein theory}

Here we obtain a metric expression for the multivielbein interactions
(\ref{mass}) before solving the constraints (\ref{Eeom-antisym}). The
difficulty encountered earlier is avoided by going away from the gauge
fixed form of (\ref{mass}) and the restricted choice of
parameterization. 

{\it Completely gauge invariant multivielbein action}\,:
In principle, when the $\cN$
metrics $g_{\mu\nu}(I)$ are written in terms of vielbeins,   
\be
\label{metrics-e} 
g_{\mu\nu}(I)=e_{~\mu}^{a}(I)\,\eta_{ab}(I)\,e^b_{~\nu}(I)\,, 
\ee 
there is no reason to identify the $\cN$ Lorentz frames with each
other and with the Lorentz frame of $\te_{a_1\hdots a_\md}$. To
emphasize this, we denote the vielbeins in the independent Lorentz
frames by $e^a_{~\mu}(I)$, in contrast to the $E^a_{~\mu}(I)$ of the
previous section. Then, in general, the Lorentz frames of the
$e^a_{~\mu}(I)$ are related to that of $\te_{a_1\hdots a_4}$ by local
Lorentz transformations (LLT) $\Lambda^{a}_{~b}(I)$, 
\be
\Lambda^{a}_{~b}(I)\,\eta^{bd}(I)\,\Lambda^{c}_{~d}(I) =\eta^{ac}(I)\,,
\label{Lambda}
\ee
and the interaction term preserving all $\cN$ local Lorentz
invariances is given by,   
\be
U_{I_1\cdots I_\md} = 
\te^{\mu_1\cdots \mu_\md}
\te_{a_1\cdots a_\md}\,\left [\Lambda^{a_1}_{~b_1}(I_1)
e^{b_1}_{~\mu_1}(I_{_1})\right]\cdots\left[\Lambda^{a_\md}_{~b_\md}(I_\md) 
e^{b_\md}_{~\mu_\md}(I_\md)\right]\,. 
\label{massLI}
\ee
The $\Lambda^{a_i}_{~b_i}(I_i)$ are St\"uckelberg fields that
restore the $\cN -1$ broken LLT's of (\ref{mass}) and transform as
bi-fundamentals. The $a_i$ index transforms under LLT of the
$\te_{a_1\cdots a_\md}$ frame, while the $b_i$ index transforms under
the LLT of the corresponding $e_{~\mu_i}^{b_i}(I_i)$ frame. 

The relation to the gauge fixed form of the interaction is easy to
see. If we choose local Lorentz gauges that set all
$\Lambda^a_{~b}(I)$ equal to each other then, since
$\det\Lambda(I)=1$, the interactions (\ref{massLI}) reduce to
(\ref{mass}), breaking the symmetry to the diagonal subgroup of the
$\cN$ LLT's. However, for generic starting $\Lambda(I)$, the vielbeins
in this gauge are now in the form (\ref{Eadm-gen}) rather than 
(\ref{Eadm}).  But, as discussed in the previous section, the no-ghost
proof of \cite{Hinterbichler:2012cn} readily extends to this case and
the completely gauge invariant form of multivielbein theory is ghost
free for general $\Lambda(I)$. To find an expression for the
multivielbein theory in terms of the metrics, one needs to choose a
different gauge.

{\it Multimetric formulation}\,: Now we consider the multivielbein
interactions in the form (\ref{massLI}). Without loss of generality,
let us pick out one of the vielbeins, say  $e_{~\mu}^{a}(1)$ and
express the volume element in terms of the corresponding metric
$g_{\mu\nu}(1)$ (\ref{metrics-e}). To do this, we first have to
identify the Lorentz frames of $e_{~\mu}^{a}(1)$ and $\te_{a_1\cdots
  a_\md}$. Since $\te_{a_1\cdots a_\md}$ is Lorentz invariant, we
have,  
\be
\te_{a_1\cdots a_\md}=\te_{c_1\cdots c_\md}(\Lambda^{-1})^{c_1}_{~a_1}(1)
\hdots (\Lambda^{-1})^{c_\md}_{~a_\md}(1) \,.
\ee 
Using this in (\ref{massLI}) gives,
\be
\te^{\mu_1\cdots \mu_\md}
\te_{c_1\cdots c_\md}\,\left [\Lambda^{c_1}_{~b_1}(1,I_{1})
e^{b_1}_{~\mu_1}(I_1)\right]\cdots\left[\Lambda^{c_\md}_{~b_\md}(1,I_\md) 
e^{b_\md}_{~\mu_\md}(I_\md)\right]\,,
\label{massLI2}
\ee
where we have defined,
\be
\Lambda^{c}_{~b}(1,I)= (\Lambda^{-1})^{c}_{~a}(1)\Lambda^{a}_{~b}(I)\,.
\ee
Obviously, $\Lambda^{c}_{~b}(1,1)=\delta^c_{~b}$. A generic
$\Lambda^{c}_{~b}(1,I)$ transforms in a bi-fundamental, the upper 
index $c$ transforming under the LLT of $e^{c}_{~\mu}(1)$ while the
lower index $b$ transforms under LLT of $e^b_{~\mu}(I)$. In this form,
the action still has all the $\cN$ local Lorentz invariances. Now we
can introduce the $g(1)$ volume element in (\ref{massLI2}) through the
identity, 
\be
\te_{c_1\cdots c_\md}=\sqrt{-\det{g(1)}}\,\te_{\nu_1\cdots \nu_\md}\, 
e^{\nu_1}_{~c_1}(1)\cdots e^{\nu_{\md}}_{~c_{\md}}(1)\,,
\label{g1ve}
\ee
that follows from the definition of the determinant.

To write the resulting expression in terms of metrics, consider (using
the obvious matrix notation $e= (e^a_{~\mu})$ for the vielbeins),  
\be
g^{-1}(1) g(I)=e^{-1}(1)\,\eta^{-1}\,e^{-1T}(1)\,e^T(I)\,\eta\,e(I)\,.
\ee
Using LLT's, $e(I)$ (for $I=2,\cdots, \cN$) can always be transformed
to $\bar e(I)$ such that the matrix $\bar e(I)e^{-1}(1)\eta^{-1}$ is
symmetric (this can be thought of as Lorentz transforming a polar
decomposition of this matrix). This condition can also be expressed
as,  
\be
\label{bmconst}
\eta^{-1} e^{-1T}(1)\bar{e}^T(I)=[\eta^{-1} e^{-1T}(1)\bar e^T(I)]^T\,.
\ee 
Then, in this local frame one has $g^{-1}(1) g(I)=[e^{-1}(1)\bar
  e(I)]^2$ or,  
\be\label{e-sym-gauge}
\bar e^b_{~\mu}(I) = e^{b}_{~\lambda}(1) \,
\left[\sqrt{g^{-1}(1) g(I)}\,\right]^\lambda_{~\mu}\,.
\ee
This also includes the trivial $I=1$ case. These gauge fixings of
$e(I)$, for $I=2\cdots \cN$, identify their local Lorentz frames
with that of $e(1)$. Now combining (\ref{massLI2}), (\ref{g1ve}),
(\ref{e-sym-gauge}) and using the notation,     
\be
L^\nu_{~\lambda}(I) \equiv
e^\nu_{~a}(1)\,\Lambda^{a}_{~b}(1,I)\,e^{b}_{~\lambda}(1)\,,
\label{ep-L}
\ee
we finally get an expression for the mass potential in terms of the
$\cN$ metrics $g(I)$,
\begin{align}
U_{I_1\cdots I_\md} = & \sqrt{-\det{g(1)}}\,\,
\te^{\mu_1\cdots \mu_\md}\,\te_{\nu_1\cdots \nu_\md}\, 
L^{\nu_1}_{~\lambda_1}(I_1)\Big[\sqrt{\ds g^{-1}(1)\,g(I_1)}\,
\Big]^{\lambda_1}_{~\mu_1}\,\cdots \nn\\ 
& \hspace{7cm} \times\, L^{\nu_\md}_{~\lambda_\md}(I_\md) 
\Big[\sqrt{\ds g^{-1}(1)\,g(I_\md)}\,\Big]^{\lambda_\md}_{~\mu_\md}\,.  
\label{massMetric}
\end{align}
Thus the multimetric interaction is a direct generalization of the
``deformed determinant'' structure \cite{Hassan:2011vm} and contains
ingredients not more complicated than the matrix square-root first 
encountered in \cite{deRham:2010kj}.  

The mass potential (\ref{massMetric}) also contains the $\cN -1$
matrices $L(I)$ (with $L^\mu_{~\nu}(1)=\delta^\mu_\nu$). In $d$
dimensions, each $L^{\nu}_{~\lambda}(I)$ (for $I>1$) inherits
$d(d-1)/2$ degrees of freedom from the local Lorentz transformation
$\Lambda^{a}_{~b}(I)$, so the pair $(\,g_{\mu\nu}(I),
L^{\nu}_{~\lambda}(I)\,)$ contains $d^2$ degrees of freedom, the same
as the content of the vielbein ${e^a}_\mu(I)$. Thus, even off-shell,
the multimetric and multivielbein forms contain the same number of
fields, after accounting for the local symmetries of the vielbein
formalism. The $d(d-1)/2$ independent components of each $L(I)$ will
be determined by their equations of motion as will be explained
below. The use of $L(I)$ makes it possible to express the mass
potential as a finite polynomial in $\sqrt{g^{-1}(1)g(I)}$.

While the vielbein formulation is convenient for some applications,
the metric formulation comes in handy when using the action to describe
interactions of spin-2 fields which, in a general covariant setup, are
described in terms of rank-2 symmetric tensors. To find the spin-2
content in the vielbein formulation one needs to eliminate $d(d-1)/2$
a priori unknown combinations of the $d^2$ components in each vielbein
by the constraint equations (\ref{Eeom-antisym}), while the metric
formulation manifestly isolates these from the symmetric tensors. 

\section{The constraint equations for \texorpdfstring{$L(I)$}{L(I)}}

The $d(d-1)/2$ independent components of each $L^\mu_{~\nu}(I)$ appear
as non-dynamical variables in the multimetric action and the
 associated equations of motion are constraints that could be solved to
determine these. This is discussed below. For the sake of familiarity,
the discussion is formulated in 4 dimensions but trivially extends
to $d$ dimensions.
 
{\it The $L(I)$ equations}\,: Let us set $g_{\mu\nu}=g_{\mu\nu}(1)$,
$e^a_{~\mu}=e^a_{~\mu}(1)$ and denote by $L(I)$ the matrix with
elements $L^\mu_{~\nu}(I)$. It satisfies the property, inherited from
(\ref{Lambda}),    
\be
L^T(I)\, g\, L(I)= g\,.
\ee
Hence $L(I)$ contains the 6 independent degrees of freedom in 
$\Lambda^{a}_{~b}$ but it also depends on $g_{\mu\nu}$. The 6
independent components are determined through their equations of
motion. To evaluate these, we first have to disentangle the Lorentz 
degrees of freedom in $L(I)$ from its dependence on $g_{\mu\nu}$.     

For this purpose we make use of the fact that Lorentz group can
be parameterized in terms of antisymmetric matrices $\hat A_{ab}$
as \footnote{This is the Cayley transform of a Lorentz transformation.
  Here it is preferred over the exponential form $e^w$, since for a
  matrix $w$, $\p(e^w)/\p w^i_j$ does not have a closed form and is
  useful only perturbatively.}
\be 
\Lambda^{a}_{~b}=\left[(\eta +\hat A)^{-1}(\eta -\hat
  A)\right]^a_{~b}\,, \qquad \hat A_{ab}=-\hat A_{ba}\,. 
\label{Cayley}
\ee 
This gives an expression for $L^\mu_{~\nu}(I)$ (\ref{ep-L}) in terms
of $A_{\mu\nu}(I)=e^a_{~\mu}e^b_{~\nu}\hat A_{ab}(I)$, 
\be 
L^\mu_{~\nu}(I)=\left[\left(\ds g+A(I)\right)^{-1}\left(g-A(I)\right)
\right]^\mu_{~\nu}\,,  
\qquad
A_{\mu\nu}(I)=-A_{\nu\mu}(I)\,. 
\ee 
Since the conditions on $A_{\mu\nu}$ do not depend on the metric, the
two can be varied independently and the 6 $A_{\mu\nu}$ equations of
motion are the needed constraints. Varying $A_{\mu\nu}$ one gets
(suppressing the $I$),  
\be
\delta L^\mu_{~\nu}= -2\left[\left(g+ A\right)^{-1}\, \delta A \,
\left(g+A\right)^{-1}
  \right]^{\mu\lambda}g_{\lambda\nu}\,, 
\ee
which, taking the antisymmetry of $A$ into account, gives
\be
\frac{\delta L^\mu_{~\nu}}{\delta A_{\rho\sigma}}=
\left[(g+A)^{-1}\right]^{\mu\rho}\left[(g+A)^{-1}\right]^{\sigma\lambda}
g_{\lambda\nu} -  
\left[(g+A)^{-1}\right]^{\mu\sigma}\left[(g+A)^{-1}\right]^{\rho\lambda} 
g_{\lambda\nu} \,.
\ee
Since the multimetric action depends on the $A(I)$ only through the
$L(I)$, one can easily obtain the equations of motion (for each
$A(I)$ and $L(I)$, $I=2,\cdots , \cN$),  
\begin{align}
\label{A-eom}
\frac{\delta U}{\delta A_{\rho\sigma}} =
\left(
\left[(g+A)^{-1}\right]^{\mu\rho}
\left[(g+A)^{-1}\right]^{\sigma\nu}-  
\left[(g+A)^{-1}\right]^{\mu\sigma}
\left[(g+A)^{-1}\right]^{\rho\nu} \right) 
\frac{\delta U}{\delta L^\mu_{~\lambda}}g_{\lambda\nu}\,=0\,.
\end{align}
These are the $6(\cN-1)$ equations that determine the independent
auxiliary fields $A_{\mu\nu}$, which, in turn, determine the relative
orientations of the Lorentz frames in (\ref{massLI}). On multiplying
by $(g-A)_{\rho\alpha} (g-A)_{\sigma\beta}$ they take the compact
form, 
\begin{align}
\label{A-eomL}
\left(L^\mu_{~\alpha}(I)\,g_{\beta\nu}-L^\mu_{~\beta}(I)\,g _{\alpha\nu}
\right) \frac{\delta U}{\delta L^\mu_{~\nu}(I)}=0\,.
\end{align}
  
From the equivalence between the metric and the vielbein formulations
it is clear that these equations must be equivalent to the $6(\cN-1)$
constraints (\ref{Eeom-antisym}) obtained in
\cite{Hinterbichler:2012cn}. To see this, note that, after 
the Lorentz frames of $e^a_{~\mu}(1)$ and $\epsilon_{a_1\cdots a_\md}$
are identified as in (\ref{massLI2}), the vielbeins $e^a_{~\mu}(I)$
are related to the $E^a_{~\mu}(I)$ of section 2 via 
\beqn
\label{relvb}
E^a_{~\mu}(I)=\Lambda^{a}_{~b}(1,I)e^b_{~\mu}(I)\,.
\eeqn
Then, converting the coordinate indices in (\ref{A-eomL}) into Lorentz  
indices by multiplication with $L_\rho^{~\alpha}(I) L_\sigma^{~\beta}(I)
{e^\rho}_{a}(1){e^\sigma}_b(1)$ and using $\frac{\delta U}
{\delta L^\mu_{~\nu}}=\frac{\delta U}{\delta
  {E^a}_\rho}\frac{\delta {E^a}_\rho}{\delta
  {\Lambda^c}_d}\frac{\delta {\Lambda^c}_d}{\delta L^\mu_{~\nu}}$
along with (\ref{relvb}) and (\ref{ep-L}), we arrive at the constraint
equations (\ref{Eeom-antisym}). In the following we exemplify the
equivalence of the constraints in the two different formulations of
the theory for two specific classes of interaction terms. 

{\it Bimetric interactions}\,: In the multivielbein formulation, for
interaction terms involving ${e^a}_\mu(1)$ and only one other
${e^a}_\mu(I)$, the constraint (\ref{Eeom-antisym}) is simply
solved by the condition (\ref{bmconst}) that expresses the vielbeins
in terms of metrics. In the multimetric formulation this translates
to the constraint (\ref{A-eomL}) having a solution $L^\mu_{~\nu}(I)
=\delta^\mu_{~\nu}$. We should therefore automatically find that (with
$L^{\mu\nu}(I)=L^\mu_{~\lambda}(I) g^{\lambda\nu}(1)$),   
\be
\label{bimetriccond}
\left.\left(\frac{\delta U}{\delta L^{\mu\nu}}\,
- \frac{\delta U}{\delta L^{\nu\mu}}\right)\right|_{A=0}=0\,.
\ee
This condition is indeed fulfilled in the metric description due to
the symmetry of, 
\be\label{mdef}
M_{\mu\nu}(I)\equiv g_{\mu\lambda}(1){\left[\sqrt{g^{-1}(1)g(I)}~
\right]^\lambda}_\nu\,,
\ee
in terms of which the mass potential in the bimetric theory has
generic terms,
\be
\sqrt{-\det{g(1)}}~\te^{\mu_1\hdots \mu_n \lambda_{n+1}\hdots\lambda_\md}\,
\te_{\nu_1\hdots \nu_n\lambda_{n+1}\hdots\lambda_\md} \,L^{\nu_1\kappa_1}(I)
M_{\kappa_1\mu_1}(I)\,
\hdots\, 
L^{\nu_n\kappa_n}(I)
M_{\kappa_n\mu_n}(I)\,.
\label{massMetric2}
\ee
Differentiating this with respect to $L^{\mu\nu}(I)$ and afterwards
setting $A(I)=0$ or $L^{\mu\nu}(I)=g^{\mu\nu}(1)$ results in a sum of
terms proportional to matrices of the form, 
\be
{M_\alpha}^{\mu_1}(I){M_{\mu_1}}^{\mu_2}(I)\hdots M_{\mu_{n-1}\beta}(I)\,.
\ee
Each of these terms is manifestly symmetric by definition of
${M^\mu}_{\nu}$, which then directly implies (\ref{bimetriccond}).
This also verifies that for $\cN=2$ one recovers the bimetric theory 
\cite{Hassan:2011zd} with the potential written in the notation of
\cite{Hassan:2011vm}.

{\it Tri-metric interactions}\,: In order to provide both an explicit
set-up in which $A=0$ is not a valid solution and another illustration
of the equivalence to the symmetry conditions arising from the
vielbein formulation, we now consider the tri-metric term
of~\cite{Hinterbichler:2012cn} that includes ${e^a}_\mu(1)$ and two
other ${e^a}_\mu(I)$, $I=2,3$. In terms of the metrics  
and with the definition (\ref{mdef}), this term reads
\be\label{triv}
\sqrt{-\det{g(1)}}~\Big(\mathrm{Tr}\Big[L(2)M(2)\Big]
\mathrm{Tr}\Big[L(3)M(3)\Big]
-\mathrm{Tr}\Big[L(2)M(2)L(3)M(3)\Big]\Big)\,.
\ee
In this case the 12 constraint equations (\ref{A-eomL}) read
\begin{align}\label{finconst}
&L^{~\nu}_{\alpha}(2)M_{\nu\beta}(2)L^{\rho\sigma}(3)M_{\rho\sigma}(3)
-L^{~\nu}_{\alpha}(2)M_{\nu\rho}(2)L^{\rho\sigma}(3)M_{\sigma\beta}(3)
-\,(\alpha\leftrightarrow\beta)=0\,,\nn\\
&L^{~\nu}_{\alpha}(3)M_{\nu\beta}(3)L^{\rho\sigma}(2)M_{\rho\sigma}(2)
-L^{~\nu}_{\alpha}(3)M_{\nu\rho}(3)L^{\rho\sigma}(2)M_{\sigma\beta}(2)
-\,(\alpha\leftrightarrow\beta)=0\,.
\end{align}
Trying to enforce these conditions for $A(I)=0$,
i.e. $L(I)=\mathbb{1}$, we see that this would require  
$M_{\sigma\alpha}(2)M^\alpha_{~\rho}(3) = M_{\rho\alpha}(2)
M^\alpha_{~\sigma}(3)$. Since the $M(I)$ generically do not commute,  
we conclude that demanding $A(I)=0$ is not compatible with the
symmetry constraints arising from the tri-vertex . 

One can also verify that the conditions (\ref{finconst}) are
equivalent to the ones derived for the tri-vertex term in
\cite{Hinterbichler:2012cn}. There, the conditions on the vielbeins
${E^a}_\mu(I)$ read, 
\begin{align}\label{vbconst}
\Big[E^{-1}(1)E(2)\eta~\mathrm{Tr}\Big(E^{-1}(1)E(3)\Big)-E^{-1}(1) 
E(2)E^{-1}(1)E(3)\eta\Big]_{ab}-(a\leftrightarrow b)&=&0\,,\nn\\
\Big[E^{-1}(1)E(3)\eta~\mathrm{Tr}\Big(E^{-1}(1)E(2)\Big)-E^{-1}(1) 
E(3)E^{-1}(1)E(2)\eta\Big]_{ab}-(a\leftrightarrow b)&=&0\,.
\end{align}
Upon substituting ${L^\mu}_L(I){M^\lambda}_\nu(I)= {E^{\mu}}_a(1) 
{E^a}_\nu(I)$ into (\ref{finconst}) and converting the Lorentz indices
to spacetime indices by multiplication with ${e^\alpha}_{a}(1)
{e^\beta}_b(1)$, we arrive at (\ref{vbconst}).

\end{document}